\documentclass[prd,twocolumn,showpacs,superscriptaddress,preprintnumbers,epsf]{revtex4}

\usepackage{graphicx}
\usepackage{dcolumn}
\usepackage{bm}
\usepackage{epsf}

\renewcommand{\bar}[1]{\overline{#1}}

\usepackage{amssymb}

\usepackage{indentfirst}
\usepackage{psfig,color}
\usepackage{epsfig}
\usepackage{epsf}
\usepackage{graphicx}

\providecommand{\Journal}[4] {#1 {\bf #2}, #3 (#4)}
\providecommand{\EPJA}{Eur. Phys. J. A } %
\providecommand{\MPLA}{Mod. Phys. Lett. A} %
\providecommand{\NPA}{Nucl. Phys. A } %
\providecommand{\NPB}{Nucl. Phys. B } %
\providecommand{\PLB}{Phys. Lett. B } %
\providecommand{\PRL}{Phys. Rev. Lett. } %
\providecommand{\PRD}{Phys. Rev. D } %
\providecommand{\PRSA}{Proc. Roy. Soc. A } %
\providecommand{\ZPA}{Z. Phys. A } %
\begin{document}

\title{The 27-plet Baryons from Chiral Soliton Models }

\author{Bin Wu}
\affiliation{Department of Physics, Peking University, Beijing
100871, China}
\author{Bo-Qiang Ma}
\altaffiliation{Corresponding author}\email{mabq@phy.pku.edu.cn}
\affiliation{ CCAST (World Laboratory), P.O.~Box 8730, Beijing 100080, China\\
Department of Physics, Peking University, Beijing 100871, China}

\begin{abstract}
We use the perturbation method to calculate the masses and widths
for 27-plet baryons with spin $\frac{3}{2}$ from chiral soliton
models. According to the masses and quantum numbers, we find all
the candidates for non-exotic members of 27-plet. The calculation
of the widths shows that these candidates manifest an approximate
symmetry of the 27 representation of the SU(3) group, and the
quantum numbers of $\Xi(1950)$ seem to be
$I(J^P)=\frac{1}{2}(\frac{3}{2}^+)$. Up to leading order of the
strange quark mass, we find that the exotic members have widths
much larger than those of the anti-decuplet members.
\end{abstract}


\vfill

\vfill

\maketitle
\par

The possible observation \cite{LEPS,DIAN,CLAS,SAPH,HERMES} of an
exotic baryon with a narrow width and a positive strangeness
number S=+1, the $\Theta^+(1540)$, has led to an explosion of
interest this year. In chiral soliton models, the anti-decuplet
baryon multiplet \cite{Mano,Chem} is the lightest one next to the
octet and decuplet, and $\Theta^+$ is the lightest member of the
anti-decuplet. In quark language, $\Theta^+$ is of the minimal
five-quark configuration $\left|uudd\bar{s}\right>$
\cite{qqqqqbar,GM99}; thus, $\Theta^+$ has the exotic strangeness
number S=+1 and, if exists, must be the lightest exotic baryon of
pentaquark states. Predictions about the mass and width of
$\Theta^+$ from chiral soliton models \cite{Pra,penta1,Diak,Weig}
have played an important role in the searches of $\Theta^+$.
Especially, in Ref.~\cite{Diak}, Diakonov, Petrov and Polyakov
predicted that $\Theta^+$ has a mass around 1.53~GeV and a width
$\Gamma_{\Theta^+}<15$~MeV, which is surprisingly close to
experimental observations. The experimental results seemed to
support chiral soliton models. Following this success, some
authors \cite{Wall,27sec} studied baryons in the 27-plet and
35-plet, which are the next baryon multiplets to the anti-decuplet
from chiral soliton models. However, a recent report \cite{NA49}
on the existence of a narrow $\Xi^-\pi^-$ baryon resonance with a
mass of $1.862\pm0.003$~GeV and width below the detector
resolution of about $0.018$~GeV, if confirmed and identified as a
member of the anti-decuplet, seems to imply that identifying the
nucleon resonance N(1710) with a member of the anti-decuplet needs
revision \cite{Diak2,Wu-Ma}.

\par

The purpose of the present note is to give a clear picture of all
the 27-plet baryons from chiral soliton models and check the
validity of this picture by symmetry. Though there are criticisms
of the validity of chiral soliton models to study pentaquark
states \cite{anti-sol}, we find that we can identify candidates
for all  non-exotic members in the 27-plet with spin 3/2,
consistent with the experimental results in \cite{NA49}. We also
make predictions about the masses and widths for all exotic
members in the 27-plet.

\par

\begin{figure}
\begin{center}
\includegraphics[width=8cm]{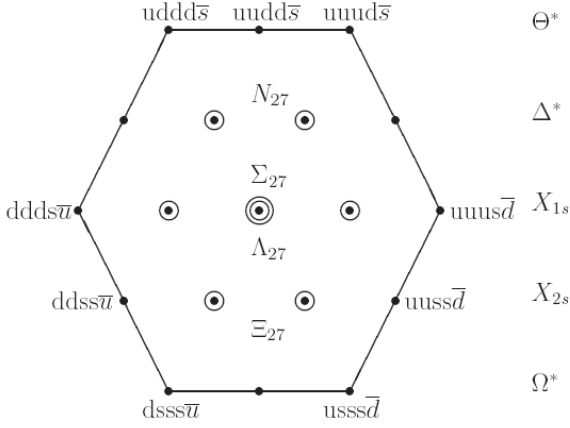}
\end{center}
\caption{The quark content of the \{27\} multiplet
baryons.}\label{fig1}
\end{figure}

The action of Skyrme model is of the form \cite{Skyr,witt}
\begin{equation}\begin{array}{rl}
I=&\frac{f^2_\pi}{4}\int d^4x\mbox{Tr}(\partial_\mu U\partial_\mu
U^\dagger)\\ &+\frac{1}{32e^2}\int d^4x\mbox{Tr}\left[\partial_\mu
UU^\dagger,\partial_\nu UU^\dagger\right]^2+N_c\Gamma,
\end{array}
\end{equation}
and the SU(3) chiral field is expressed as
\begin{equation}
    U(x)=\exp{\left[i\frac{\lambda_b\phi_b(x)}{f_\pi}\right]}=A(t)U_1(\mathbf{x})A(t)^{-1},\ \  A \in SU(3),
\end{equation}
where $f_\pi\approx 93$~MeV is the observed pion decay constant,
the dimensionless parameter $e$ is introduced to stabilize the
solitons by Skyrme, $\Gamma$ is the Wess-Zumino term, $\lambda_b$
are the eight Gell-Mann SU(3) matrices, $\phi_b(x)$ are the eight
pseudoscalar meson fields, and $U_1(\mathbf{x})$ is a solitonic
solution (with unit baryonic charge) of the equation of motion
\cite{Guad}
\begin{equation}
    U_1(\mathbf{x})=\left(
        \begin{array}{cc}
            \exp{[i(\mathbf{\widehat{r}}\cdot\mathbf{\tau})F(r)]} & \begin{array}{c}0\\0\end{array}\\
            \begin{array}{cc}0&0\end{array}&1
        \end{array}
        \right),
\end{equation}
where $F(r)$ is the spherical-symmetric profile of the soliton,
$\mathbf{\tau}$ are the three Pauli matrices, and
$\mathbf{\widehat{r}}$ is the unit vector in space. The
eigenvalues of the collective Hamiltonian are \cite{Guad}
\begin{equation}
\begin{array}{lr}
E^{(p,q)}_J=&M_{cl}+\frac{1}{6I_2}\left[p^2+q^2+pq+3(p+q)-\frac{1}{4}(N_cB)^2\right]\\
&+\left(\frac{1}{2I_1}-\frac{1}{2I_2}\right)J(J+1)\label{H},
\end{array}
\end{equation}
where $(p,q)$ denotes an irreducible representation of the SU(3)
group, $M_{cl}$, $I_1$ and $I_2$ are given by the 3-dimensional
space coordinate integrals of even functions of $F(r)$ and $e$,
treated model-independently and fixed by experimental data,
$M_{cl}$ is the classical soliton mass, $I_1$ and $I_2$ are
moments of inertia. From the energy eigenvalues above, it can be
seen that \{27\} multiplet with spin 3/2 is next to the
anti-decuplet, whose quark content is suggested in
Fig.~\ref{fig1}.

\par

Using the wave function $\Psi_{\nu\nu^\prime}^{(\mu)}$ of baryon
$B$ in the collective coordinates
\begin{equation}
    \Psi_{\nu\nu^\prime}^{(\mu)}(A)=\sqrt{\mbox{dim}(\mu)}D^{(\mu)}_{\nu\nu^\prime}(A)\label{psi},
\end{equation}
where $(\mu)$ denotes an irreducible representation of the SU(3)
group; $\nu$ and $\nu^{\prime}$ denote $(Y, I, I_3)$ and $(1, J,
-J_3)$ quantum numbers collectively; $Y$ is the hypercharge of
$B$; $I$ and $I_3$ are the isospin and its third component of $B$
respectively; $J_3$ is the third component of spin $J$; and
$D^{(\mu)}_{\nu\nu^\prime}(A)$ are representation matrices, we can
deal with the symmetry breaking Hamiltonian \cite{Bolt}
perturbatively \cite{Park}
\begin{eqnarray}
    &&H^\prime=\alpha D^{(8)}_{88}+\beta
    Y+\frac{\gamma}{\sqrt{3}}\sum_{i=1}\limits^3D^{(8)}_{8i}J^i\label{Hp},
\end{eqnarray}
where the coefficients $\alpha$, $\beta$, $\gamma$ are
proportional to the strange quark mass and model dependent. In
this note they are treated model-independently and fixed by
experiments. $D^{(8)}_{ma}(A
)=\frac{1}{2}\mbox{Tr}(A^{\dagger}\lambda^mA\lambda^a)$ is the
adjoint representation of the SU(3) group.

We can use the relations between the masses of the octet and
decuplet baryons, then we need two additional equations to fix the
parameters. Up to know, we have two methods to fix all the
parameters in (\ref{H}) and (\ref{Hp}) model-independently: (I)
take $\Theta^+$ as the member of the anti-decuplet, and use
relations about the mass (1.54~GeV) and width ($<$25~MeV) of
$\Theta^+$; (II) take both $\Theta^+$ and the candidate for
$\Xi_{3/2}$ \cite{NA49} as members of the anti-decuplet and only
use the mass relations of the anti-decuplet baryons
\cite{Diak2,Wu-Ma}. For method I, we have the results
\[
\begin{array}{lll}
\alpha=-766~\mbox{MeV},&
\beta=22~\mbox{MeV},&\gamma=254~\mbox{MeV},\\
1/I_1=154~\mbox{MeV},&1/I_2=376~\mbox{MeV};
\end{array}\]
and for method II we have the results
\[
\begin{array}{lll}
\alpha=-663~\mbox{MeV},&
\beta=-12~\mbox{MeV},&\gamma=185~\mbox{MeV},\\
1/I_1=154~\mbox{MeV},&1/I_2=399~\mbox{MeV}.
\end{array}\]
The predicted mass of $\Xi_{3/2}$ from method I is 1.81,  which is
compatible with the experimental observation \cite{NA49} and can
be further adjusted to meet the data within uncertainties.

We find that the masses of the 27-plet calculated by the two
methods are nearly equal, shown in Table~1. These results are
close to those calculated by Walliser and Kopeliovich \cite{Wall},
with difference in the mass of $\Delta_{27}$. This shows the
validity of the use of the perturbation method in chiral soliton
models. In Table~1, we also list the candidates for the 27-plet
baryons. We can find all the candiates for the non-exotic members
by considering their masses and $I(J^{P})$ in the baryon listing
\cite{PDG}. To verify this identification, we calculate the widths
of the 27-plet baryons.
\begin{widetext}\begin{center}
\begin{tabular}{|l|l|l|l|l|l|l|l|}
\multicolumn{7}{c}{Table 1. The masses (GeV) of baryons in the
\{27\} multiplet}
\\ \hline&$\left<B|H^\prime|B\right>$&
method I&method II&candidate&$I(J^{P})$&PDG\\
\hline
$\Delta^{*}$&$\frac{13}{112}\alpha+\beta-\frac{65}{224}\gamma$&1.62&1.64
&$\Delta(1600)$&$\frac{3}{2}(\frac{3}{2}^+)$&$1.55~\mbox{to}~1.70$\\
\hline
$N_{27}$&$\frac{1}{28}\alpha+\beta-\frac{5}{56}\gamma$&1.73&1.73&N(1720)
&$\frac{1}{2}(\frac{3}{2}^+)$&$1.65~\mbox{to}~1.75$\\
$\Sigma_{27}$&$-\frac{1}{56}\alpha+\frac{5}{112}\gamma$&1.79&1.80
&$\Sigma(1840)$&$1(\frac{3}{2}^+)$&$1.72~\mbox{to}~1.93$\\
$\Xi_{27}$&$-\frac{17}{112}\alpha-\beta+\frac{85}{224}\gamma$&1.95&1.96
&$\Xi(1950)$&$\frac{1}{2}(\frac{3}{2}^+)(?^?)$&$1.95\pm0.015$\\
$\Lambda_{27}$&$-\frac{1}{14}\alpha+\frac{5}{28}\gamma$&1.86&1.86&$\Lambda(1890)$&$0(\frac{3}{2}^+)$&$1.85~\mbox{to}~1.91$\\
\hline
$\Theta^{*}$&$\frac{\alpha}{7}+2\beta-\frac{5}{14}\gamma$&1.61&1.60&?&$1(\frac{3}{2}^+)?(?^?)$&?\\
$X_{1s}$&$\frac{5}{56}\alpha-\frac{25}{112}\gamma$&1.64&1.68&?&$2(\frac{3}{2}^+)?(?^?)$&?\\
$X_{2s}$&$-\frac{1}{14}\alpha-\beta+\frac{5}{28}\gamma$&1.84&1.87&?&$\frac{3}{2}(\frac{3}{2}^+)?(?^?)$&?\\
$\Omega^{*}$&$-\frac{13}{56}\alpha-2\beta+\frac{65}{112}\gamma$&2.06&2.07&?&$1(\frac{3}{2}^+)?(?^?)$&?\\
\hline
\end{tabular}
\end{center}\end{widetext}

The decay of a 27-plet baryon $B$ to an octet baryon $B^{\prime}$
and a pseudoscalar meson $m$ is controlled by a pseudoscalar
Yucawa coupling~\cite{Blot1,Diak}:
\begin{equation}
    \widehat{g}_A\propto~G_0D^{(8)}_{m3}
    -G_1d_{3ab}D^{(8)}_{ma}J_b
    -\frac{G_2}{\sqrt{3}}D^{(8)}_{m8}J_3,
\end{equation}
where $d_{iab}$ is the SU(3) symmetric tensor, $a, b=4,5,6,7$, and
$J_a$ are the generators of the infinitesimal SU$_R$(3) rotations.
$G_{1}$, $G_{2}$ are dimensionless constants, $1/N_c$ suppressed
relative to $G_0$. $G_2$ is neglected, then $G_0$ and $G_1$ can be
fixed by experiments. Up to leading order of the strange quark
mass, we have
\begin{widetext}\begin{center}
\begin{equation}
    \Gamma(B\rightarrow B^\prime m)=\frac{G_s^2}{4\pi}\frac{|\mathbf{p}|}{m_B}
    \left[(m_{B^\prime}^2+\mathbf{p}^2)^{\frac{1}{2}}-m_{B^\prime}\right]\left\{\begin{array}{c}
    \frac{\mbox{dim}(\mu^\prime)}{\mbox{dim}(\mu)}\left|\begin{array}{c}\sum\limits_\gamma\left(
    \begin{array}{cc}8&\mu^\prime\\ Y_mI_m&Y_\rho I_\rho \end{array}\right|
    \left.\begin{array}{c}\mu_\gamma\\Y_\nu
    I_\nu\end{array}\right)
    \left(
    \begin{array}{cc}8&\mu^\prime\\01&1 J_\rho \end{array}\right|
    \left.\begin{array}{c}\mu_\gamma\\1J_\nu\end{array}\right)\end{array}\right|^2\end{array}\right\},
\end{equation}
\end{center}\end{widetext}
where we postulate $B$ with $(Y, I, I_3; J^P, -J_3)=(Y_\nu, I_\nu,
I_{\nu3}; J_\nu^+, -J_{\nu3})$, $B^\prime$ with $(Y, I, I_3; J^P,
-J_3)=(Y_\rho, I_\rho, I_{\rho3}; J_\rho^+, -J_{\rho3})$ and $m$
with $(Y, I, I_3; J^P, -J_3)=(Y_m, I_m, I_{m3}; 0^-, 0)$; and
$G_s^2=3.84(G_0-\frac{1}{2}G_1)^2$. If we postulate the width of
$\Theta^+~$ $\Gamma_{\Theta^+}<25~\mbox{MeV}$, we can calculate
the upper bounds of widths for all the 27-plet baryons, listed in
Table~2. We can see that the candidates for non-exotic baryons
manifest the approximate symmetry of the 27 representation of the
SU(3) group. In the results above, we only consider the flavor
SU(3) as an exact symmetry. If we take into account the effects of
flavor asymmetry, the width of $\Theta^*$ will fall by about 30\%
\cite{Wu-Ma-plb}.

\begin{widetext}
\begin{center}
\begin{tabular}{|l|l|l|l|l|l|}
\multicolumn{6}{c}{Table 2. The widths (MeV) of baryons in the
27-plet}\\\hline & PDG estimation& modes&branching
ratios&$\Gamma_i$ from data&width $\leq$calculation
\\\hline $\Delta(1600)$&$250~\mbox{to}~450$&$N\pi$&$10~\mbox{to}~25\%$&$25~\mbox{to}~113$&130\\\hline
N(1720)&$100~\mbox{to}~200$&$N\pi$&$10~\mbox{to}~20\%$&$10~\mbox{to}~40$&19\\
&&$N\eta$&$(4.0\pm1.0)\%$&$3~\mbox{to}~10$&66\\
&&$\Lambda K$&$1~\mbox{to}~15\%$&$1~\mbox{to}~30$&18\\
&&$\Sigma K$&&&0.39\\\hline
$\Sigma(1840)$&$65~\mbox{to}~120$&$N\bar{K}$&$0.37\pm 0.13$&$11~\mbox{to}~60$&50\\
&&$\Lambda\pi$&&&0\\\hline
$\Lambda(1890)$&$60~\mbox{to}~200$&$N\bar{K}$&$20~\mbox{to}~35\%$&$12~\mbox{to}~70$&46\\
&&$\Sigma\pi$&$3~\mbox{to}~10\%$&$2~\mbox{to}~30$&5\\\hline
$\Xi(1950)$&$60\pm 20$&$\Lambda\bar{K}$&seen&&90\\
&&$\Sigma\bar{K}$&possibly seen&&6.5\\
&&$\Xi\pi$&seen&&8.3\\\hline $\Theta^{*}$&?&$KN$&?&?&79\\\hline
$X_{1s}$&?&$\Sigma\pi$&?&?&96\\\hline
$X_{2s}$&?&$\Xi\pi$&?&?&58\\
&&$\Sigma\bar{K}$&?&?&36\\\hline
$\Omega^{*}$&?&$\Xi\bar{K}$&?&?&107\\
\hline
\end{tabular}
\end{center}
\end{widetext}

In summary, we use the perturbation method to deal with the
27-plet baryons with spin 3/2 from chiral soliton models.
Calculations of the widths of the candidates for the non-exotic
members manifest an approximate symmetry of the 27 representation
of the SU(3) group. Thus, it seems that chiral soliton models are
able to give us a clear picture of the 27-plet with spin 3/2, as
well as the anti-decuplet \cite{Diak2,Wu-Ma}, beyond their
validity of describing the octet and decuplet baryons. We also
predict the masses and widths of the exotic members in 27-plet.
The exotic members seem to be more difficult to be found
experimentally for their larger widths compared with those of the
anti-decuplet members. If this picture is right, the non-exotic
member $\Xi(1950)$ should be with $J^P=\frac{3}{2}^+$.

We are grateful for discussions with Yanjun Mao. This work is
partially supported by National Natural Science Foundation of
China under Grant Numbers 10025523 and 90103007.



\begin{thebibliography}{99}
\bibitem{LEPS}
T.~Nakano {\it et al.}, \Journal{\PRL} {91}{012002}{2003}.

\bibitem{DIAN}
DIANA Collaboration, V.V.~Barmin {\it et al.}, 
Phys. Atom. Nucl. {\bf 66}, 1715 (2003).

\bibitem{CLAS}
CLAS Collaboration, S.~Stepanyan {\it et al.}, 
\Journal{\PRL}{91}{252001}{2003}; V.~Kubarovsky {\it et al.},
hep-ex/0311046.

\bibitem{SAPH}
SAPHIR Collaboration, J.~Barth {\it et al.}, 
\Journal{\PLB}{572}{127}{2003}.

\bibitem{HERMES}
HERMES Collaboration, A.~Airapetian {\it et al.}, hep-ex/0312044.

\bibitem{Mano}
A.V.~Manohar, \Journal{\NPB} {248}{19}{1984}.

\bibitem{Chem}
M.~Chemtob, \Journal{\NPB} {256}{600} {1985}.

\bibitem{qqqqqbar}
R.L.~Jaffe, 
Proc. Topical Conference on Baryon Resonances, Oxford, July 1976,
SLAC-PUB-1774.

\bibitem{GM99}
H.~Gao and B.-Q.~Ma, \Journal{\MPLA} {14}{2313}{1999}.

\bibitem{Pra}
M.~Praszalowicz, in {\it Skyrmions and Anomalies} ( M.~Jezabek and
M.~Praszalowicz, eds.), World Scientific (1987), 112-131;

M.~Praszalowicz, 
\Journal{\PLB}{575}{234}{2003}.

\bibitem{penta1}
H.~Walliser, \Journal{\NPA} {548}{649}{1992}.

\bibitem{Diak}
D.~Diakonov, V.~Petrov, and M.~Polyakov, \Journal{\ZPA}
{359}{305}{1997}.

\bibitem{Weig}
H.~Weigel, \Journal{\EPJA} {2}{391}{1998}.

\bibitem{Wall}H.~Walliser and V.B.~Kopeliovich,  \Journal{J. Exp. Theor. Phys.} {97}{433}{2003}.

\bibitem{27sec}
D.~Borisyuk, M.~Faber, and
A.~Kobushkin, hep-ph/0307370.

\bibitem{NA49}
NA49 Collaboration, C.~Alt {\it et al.}, hep-ex/0310014.

\bibitem{Diak2}
D.~Diakonov and V.~Petrov, hep-ph/0310212.

\bibitem{Wu-Ma}
B.~Wu and B.-Q.~Ma, hep-ph/0311331.


\bibitem{anti-sol}
T.D.~Cohen, hep-ph/0309111; N.~Itzhaki {\it et al.},
hep-ph/0309305;

\bibitem{Skyr}
T.H.R.~Skyrme, \Journal{\PRSA} {260}{127}{1961}.

\bibitem{witt}
E.~Witten, \Journal{\NPB} {223}{422 and 433}{1983}.

\bibitem{Guad}
E.~Guadagnini, \Journal{\NPB} {236}{35}{1984}.

\bibitem{Bolt}A.~Blotz {\it et al.}, \Journal{\PLB} {287}{29}{1992};
\Journal{\NPA}{555}{765}{1993}.

\bibitem{Park}
N.W.~Park, J.~Schechter, and H.~Weigel,
\Journal{\PLB}{224}{171}{1989}.

\bibitem{PDG}Particle Data Group (PDG), K.~Hagiwara {\it et al.}, \Journal{\PRD} {66}{010001}{2002}.

\bibitem{Blot1}A.~Blotz, M.~Praszalowicz, and
K.~Goeke, \Journal{\PRD} {53}{485}{1996}.

\bibitem{Wu-Ma-plb}
B.~Wu and B.-Q.~Ma, hep-ph/0312326, Phys. Lett. {\bf B}, in press.

\end{thebibliography}
\end{document}